# Clinica: an open-source software platform for reproducible clinical neuroscience studies

Alexandre Routier[1,2,3,4,5,6], Ninon Burgos[2,3,4,5,6,1], Mauricio Díaz[6], Michael Bacci[1,2,3,4,5,6], Simona Bottani[1,2,3,4,5,6], Omar El-Rifai[1,2,3,4,5,6], Sabrina Fontanella[2,3,4,5,6,1], Pietro Gori[1,2,3,4,5,6], Jérémy Guillon[2,3,4,5,6,1], Alexis Guyot[2,3,4,5,6,1], Ravi Hassanaly[2,3,4,5,6,1], Thomas Jacquemont[2,3,4,5,6,1], Pascal Lu[2,3,4,5,6,1], Arnaud Marcoux[2,3,4,5,6,1], Tristan Moreau[2,3,4,5,6,8], Jorge Samper-González[2,3,4,5,6,1], Marc Teichmann[2,3,4,5,6,8,9], Elina Thibeau-Sutre[2,3,4,5,6,1], Ghislain Vaillant[2,3,4,5,6,1], Junhao Wen[2,3,4,5,6,1], Adam Wild[2,3,4,5,6,1], Marie-Odile Habert[10,11,12], Stanley Durrleman[1,2,3,4,5,6], Olivier Colliot[2,3,4,5,6,1,*]

[1] Inria, Aramis project-team, Paris, France

[2] Sorbonne Université, Paris, France

[3] Institut du Cerveau – Paris Brain Institute - ICM, Paris, France

[4] Inserm, Paris, France

[5] CNRS, Paris, France

[6] AP-HP, Hôpital de la Pitié-Salpêtrière, Paris, France

[7] Inria Paris, SED, Paris, France

[8] Institut du Cerveau – Paris Brain Institute - ICM, Paris, France

[9] Department of Neurology, Institute for Memory and Alzheimer's Disease, Pitié-Salpêtrière Hospital, AP-HP, Paris, France

[10] Sorbonne Université, CNRS, INSERM, Laboratoire d'Imagerie Biomédicale, LIB, Paris, France

[11] AP-HP, Hôpital Pitié-Salpêtrière, Médecine Nucléaire, Paris, France

[12] Centre d'Acquisition et Traitement des Images (CATI), www.cati-neuroimaging.com

**\* Correspondence:**
Olivier Colliot, PhD – olivier.colliot@sorbonne-universite.fr

ICM – Paris Brain Institute

ARAMIS team Pitié-Salpêtrière Hospital

47-83, boulevard de l'Hôpital, 75651 Paris Cedex 13, France

**Abstract**

We present Clinica (www.clinica.run), an open-source software platform designed to make clinical neuroscience studies easier and more reproducible. Clinica aims for researchers to i) spend less time on data management and processing, ii) perform reproducible evaluations of their methods, and iii) easily share data and results within their institution and with external collaborators. The core of Clinica is a set of automatic pipelines for processing and analysis of multimodal neuroimaging data (currently, T1-weighted MRI, diffusion MRI and PET data), as well as tools for statistics, machine learning and deep learning. It relies on the brain imaging data structure (BIDS) for the organization of raw neuroimaging datasets and on established tools written by the community to build its pipelines. It also provides converters of public neuroimaging datasets to BIDS (currently ADNI, AIBL, OASIS and NIFD). Processed data include image-valued scalar fields (e.g. tissue probability maps), meshes, surface-based scalar fields (e.g. cortical thickness maps) or scalar outputs (e.g. regional averages). These data follow the ClinicA Processed Structure (CAPS) format which shares the same philosophy as BIDS. Consistent organization of raw and processed neuroimaging files facilitates the execution of single pipelines and of sequences of pipelines, as well as the integration of processed data into statistics or machine learning frameworks. The target audience of Clinica is neuroscientists or clinicians conducting clinical neuroscience studies involving multimodal imaging, and researchers developing advanced machine learning algorithms applied to neuroimaging data.

# 1 Introduction

Neuroimaging plays an important role in clinical neuroscience studies. While the meaning of clinical neuroscience studies may vary, we use it to refer to studies involving human participants (i.e. patients with neurological and psychiatric diseases, and control subjects) explored with multimodal data (neuroimaging, clinical and cognitive evaluations, genetic data...) and most often involving longitudinal follow-up. Carrying out such studies involves many data analysis steps including image pre-processing, extraction of image-derived measurements and statistical analysis, thus requiring a wide range of expertise. A similar situation is faced by researchers in machine learning for neuroimaging: various steps are needed to extract features that are then fed to advanced learning algorithms.

The first issue met when working on clinical studies concerns the organization of neuroimaging datasets within or between institutions. The lack of a consistent structure makes arduous the sharing or reuse of data. This is true for in-house, but also for publicly available neuroimaging datasets. Another consequence of the lack of standard is the difficulty to apply automatic pipelines (e.g. extraction of neuroimaging features, statistical analysis or machine learning) and to perform quality assurance. The second issue faced by researchers processing data from clinical studies is related to the high number of software packages, such as FreeSurfer[1] (Fischl, 2012), FMRIB Software Library (FSL)[2] (Jenkinson et al., 2012) or Statistical Parametric Mapping[3] (SPM) (Friston et al., 2007), that exist in the community. Researchers have to understand the methodology behind each tool (e.g. segmentation, registration, etc.) and master them from a programming perspective before being able to combine them and develop image processing pipelines. Moreover, such "handicraft" approach makes it difficult to transmit tools and knowledge, and to merge and share results of several studies due to the heterogeneous organization of outputs. Finally, the difficulty to access or share both raw and processed neuroimaging data hinders the reproducibility of neuroimaging studies (Poline et al., 2012).

Major progress has been made in the last years to ease neuroimaging studies. First, difficulties related to the heterogeneity of image processing tools have been partly handled by the Nipype (Neuroimaging in Python – Pipelines and Interfaces) software package[4] (Gorgolewski et al., 2011). Nipype is an open-source Python project that provides a uniform environment facilitating interaction between neuroimaging software tools or algorithms, regardless of their programming language, within a single workflow. Later, the issues related to the organization of the clinical and imaging data have been tackled by the Brain Imaging Data Structure (BIDS) (Gorgolewski et al., 2016), a new standard from the community for the community. The BIDS standard is based on a file hierarchy rather than on a database management system, thus facilitating its deployment in any environment. Thanks to its clear and simple way to describe neuroimaging and behavioral data, the BIDS standard has been easily adopted by the neuroimaging community. Organizing a dataset following the BIDS hierarchy simplifies the execution of neuroimaging software tools, resulting in the development of user-friendly software. For instance, BIDS Apps (Gorgolewski et al., 2017) provides a set of pipelines for the processing of neuroimaging data that follow a BIDS hierarchy. Currently, it mainly wraps neuroimaging software packages from the community into a Docker image and is used via a command line interface. Moreover, the Nilearn[5] (Abraham et al., 2014) package facilitates the application of advanced machine learning approaches to neuroimaging data. To that purpose, it leverages the scikit-learn library[6] (Pedregosa et al., 2011) and provides tools for handling and visualizing different types of neuroimaging data and building predictive models.

---

[1] http://surfer.nmr.mgh.harvard.edu/

[2] https://fsl.fmrib.ox.ac.uk/

[3] http://www.fil.ion.ucl.ac.uk/spm/

[4] https://nipype.readthedocs.io

Nevertheless, carrying out a multimodal neuroimaging study remains challenging due to the know-how necessary to grasp each modality and tool involved. While technical implementation has been facilitated by Nipype, the development of a pipeline still requires substantial programming skills and time to master both the neuroimaging software tools and Nipype. While the BIDS standard is being adopted by the scientific community, not all public neuroimaging datasets provide a BIDS version of their data. Besides, performing a single or multimodal neuroimaging study will also require methodological expertise. For instance, a classification study of healthy subjects and patients with a neurodegenerative disease using $^{18}$F-fluorodeoxyglucose positron emission tomography (FDG PET) could involve notions of multimodal registration between FDG PET and T1-weighted magnetic resonance imaging (MRI), tissue segmentation of T1-weighted (T1w) MRI, PET partial volume correction, normalization into a standard space, and machine learning-based classification, as well as know-how of the tools used to perform these steps. Moreover, the image processing steps need to be chained from one to the other and the absence of data organization for processed neuroimages makes data analysis more complex. Finally, the neuroimaging features generated by the pipelines need to be correctly connected to statistical or machine learning frameworks.

Clinica (www.clinica.run) aims to make clinical research studies easier and pursues the community effort of reproducibility. The core of Clinica is a set of automatic pipelines for processing and analysis of multimodal neuroimaging data (currently, T1w MRI, diffusion MRI and PET data), as well as tools for statistics, machine learning and deep learning. Clinica relies on tools written by the scientific community and provides converters of public neuroimaging datasets to BIDS, processing pipelines and organization for processed files, statistical analysis, and machine learning algorithms.

The target audience is mainly of two types. First, neuroscientists or clinicians conducting clinical neuroscience studies involving multimodal imaging, typically not experts in image processing for all of the involved imaging modalities or in statistical analysis. They will benefit from a unified set of tools covering the complete set of steps involved in a study (from raw data to statistical analysis). Second, researchers developing advanced machine learning algorithms, typically not experts in brain image analysis. They will benefit from tools to convert public datasets into BIDS, fully automatic feature extraction methods, and baseline classification algorithms to which they could compare their results. Overall, we hope that Clinica will allow users to spend less time on data management and processing, to perform reproducible evaluations of their methods, and to easily share data and results within their institution and with external collaborators.

---

[5] https://nilearn.github.io

[6] https://scikit-learn.org/

## 2    Clinica overview

Clinica is an open-source software platform for reproducible clinical neuroimaging studies. It can take as inputs different neuroimaging modalities, currently anatomical MRI, diffusion MRI and PET. Clinica provides processing pipelines that involve the combination of different software packages. It currently relies on FreeSurfer (Fischl, 2012), FSL (Jenkinson et al., 2012), SPM (Frackowiak et al., 1997), Advanced Normalization Tools (ANTs)[7] (Avants et al., 2014a), MRtrix3[8] (Tournier et al., 2012), and the PET Partial Volume Correction (PETPVC) toolbox[9] (Thomas et al., 2016). The pipelines are written using Nipype (Gorgolewski et al., 2011). Features extracted with the different pipelines can be used as inputs to statistical analysis, which relies on SPM (Frackowiak et al., 1997) and SurfStat[10] (Worsley et al., 2009), or machine learning analysis, which relies on scikit-learn (Pedregosa et al., 2011) and PyTorch (Paszke et al., 2019).

Input neuroimaging data are expected to follow the BIDS data structure (Gorgolewski et al., 2016), as explained in section 3.2. Since this new standard has only recently been adopted by the community, not all public neuroimaging datasets are yet proposed in BIDS format. To facilitate the adoption of BIDS, Clinica curates several publicly available neuroimaging datasets and provides tools to convert them into the BIDS format. Processed data are organized following the ClinicA Processed Structure (CAPS) format, detailed in section 3.3, which shares the same philosophy as BIDS. Finally, a set of tools is provided to handle input and output data generated by Clinica, thus facilitating data management or connection to statistical or machine learning analysis.

A schematic overview of Clinica can be found in Figure 1. The list of pipelines currently available in Clinica is presented in Figure 2. The main functionalities of Clinica are described in the paper, but for further details the reader can refer to the documentation available on the website[11]. For each pipeline, the reader will find a description of its functionalities, a list of the tools on which it relies, an example showing how to run the pipeline, and a description of the outputs generated. The documentation of a pipeline can have several levels of reading, which are respectively targeting people new or familiar with neuroimaging and scientists working on pattern recognition and machine learning. User support is handled through a forum[12] as well as using the issue tracker on GitHub[13].

---

[7] https://stnava.github.io/ANTs/

[8] http://mrtrix.org

[9] https://github.com/UCL/PETPVC

[10] http://www.math.mcgill.ca/keith/surfstat/

[11] http://www.clinica.run/doc

[12] https://groups.google.com/g/clinica-user

[13] https://github.com/aramis-lab/clinica/issues

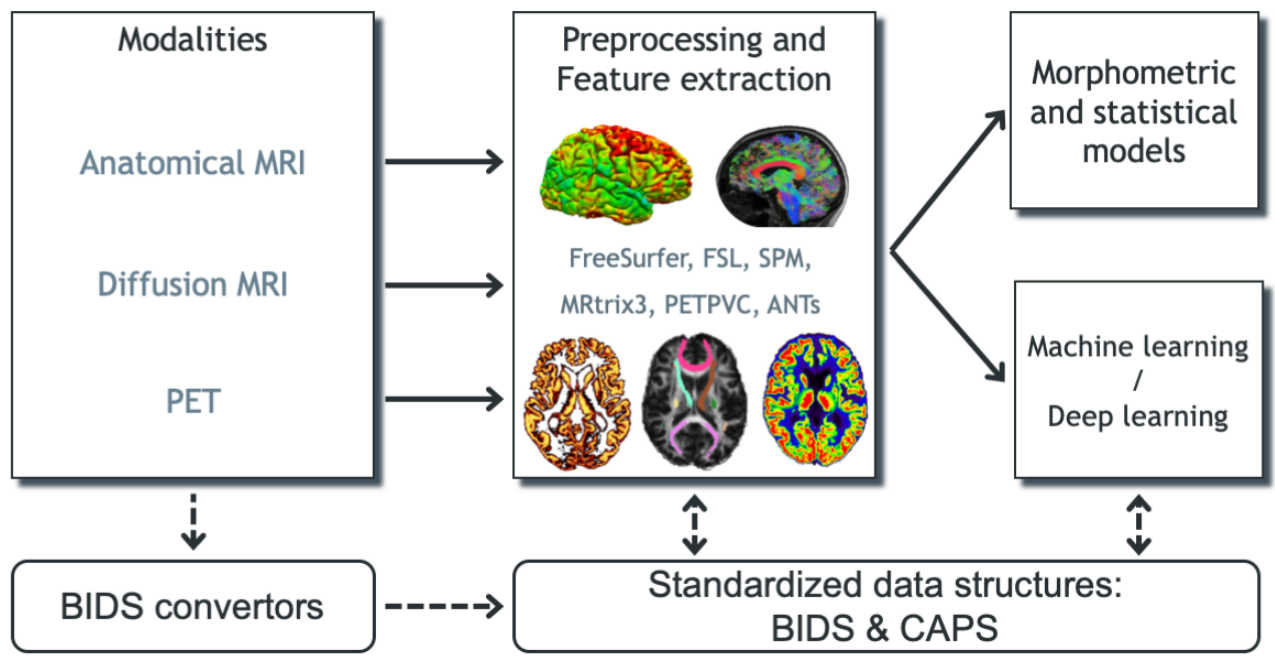

**Figure 1.** Overview of Clinica's functionalities. Clinica provides processing pipelines for MRI and PET images that involve the combination of different software packages, and whose outputs can be used for statistical or machine learning analysis. Clinica expects data to follow the Brain Imaging Data Structure (BIDS) and provides tools to convert public neuroimaging datasets into the BIDS format. Output data are stored using the ClinicA Processed Structure (CAPS).

## 3 Clinica environment

### 3.1 Software architecture of Clinica

The core of Clinica is written in Python and mainly relies on the Nipype framework (Gorgolewski et al., 2011) to create pipelines. Python dependencies also include NumPy (van der Walt et al., 2011), NiBabel (Brett et al., 2019), Pandas (McKinney, 2010), NIPY, SciPy (Jones et al., 2001), scikit-learn (Pedregosa et al., 2011), scikit-image (van der Walt et al., 2011), nilearn (Abraham et al., 2014) and PyTorch (Paszke et al., 2019).

Clinica is provided to the end user in the form of a Python package distributed through Python Package Index (PyPI) and can simply be installed by typing `pip install clinica` through the terminal, within a virtual environment.

The main usage of Clinica is through the command line, which is facilitated by the support of autocompletion. The commands are gathered into four main categories. The first category of command line (`clinica run`) allows the user to run the different pipelines on neuroimaging datasets following a BIDS or CAPS hierarchy. The `clinica convert` category allows the conversion of publicly available neuroimaging datasets into a BIDS hierarchy. To help with data management, the `clinica iotools` category comprises a set of tools that allows the user to handle BIDS and CAPS datasets, including generating lists of subjects or merging all tabular data into a single TSV file for analysis with external statistical software packages. Finally, the last category (`clinica generate`) is dedicated to developers and currently generates the skeleton for a new pipeline. Examples of command line can be found in Table 1.

**Table 1.** Categories of command line

| Category of command line | Description | Example usage |
|---|---|---|
| `run` | Run pipelines on BIDS or CAPS datasets.<br><br>The full list of pipelines is available in **section 4**. | `clinica run t1-linear bids_directory caps_directory` |
| `convert` | Convert public neuroimaging datasets into BIDS.<br><br>The list of the public datasets can be found in **section 5.1**. | `clinica convert adni-to-bids dataset_directory clinical_data_directory bids_directory` |
| `iotools` | Set of tools to handle BIDS and CAPS datasets.<br><br>The list of the I/O tools can be found in **section 5.2**. | `clinica iotools merge-tsv bids_directory --caps_directory caps_directory my_population.tsv` |
| `generate` | (For developers) Generate the skeleton source code for a new pipeline. | `clinica generate template "Modality Feature Extracted" pipeline_folder` |

## 3.2 Input data with the BIDS standard

When dealing with multiple datasets, it is difficult to automate the execution of neuroimaging pipelines since their organization may vary from each other or even within each individual dataset. If we consider neuroimaging datasets involving many participants, the lack of a clear structure will necessitate a large amount of time to curate these databases and make them easily usable. Besides, large databases are often associated with database management systems, which involve additional technical and financial resources to be maintained.

BIDS (Gorgolewski et al., 2016) is a community standard enabling the storage of multiple neuroimaging modalities and behavioral data. The BIDS standard provides a unified structure and makes easier the development and distribution of code that uses neuroimaging datasets. Moreover, the BIDS format is based on a file hierarchy rather than on a database management system, thus avoiding the installation and maintenance of additional software. As a result, BIDS can be easily deployed in any environment. The specification is intentionally based on simple file formats and folder structures to reflect current laboratory practices, which makes it accessible to a wide range of scientists coming from different backgrounds. People unfamiliar with the BIDS format can see an example of a BIDS folder in Figure 3.

For these reasons, we also adopted this standard and Clinica expects that the input data are BIDS-compliant for the execution of pipelines. Note that if a cross-sectional dataset (i.e. with no `session` folder) is provided, Clinica will interactively propose to convert the cross-sectional dataset into a longitudinal dataset with a unique session.

### 3.3   Input/Output data with the CAPS structure

Clinica has its own specifications for storing processed data, called CAPS (ClinicA Processed Structure). Of note, there exists an ongoing initiative called BIDS-derivatives that aims to provide a BIDS standard for processed data. However, we wrote the CAPS specification before the start of the BIDS-derivatives which explains why Clinica does not use the latter. Moreover, in their current state, several outputs needed by Clinica are not covered or well adapted. In particular, the notion of group does not exist yet. Nonetheless, we made humble contributions to BIDS-derivatives and we aim to increasingly contribute. Ultimately, the two specifications will probably converge.

Processed data include image-valued scalar fields (e.g. segmentation labels, tissue maps), meshes, mesh-valued scalar fields (e.g. cortical thickness maps), deformation fields, scalar outputs (e.g. volumes, regional averages), etc. Carrying out a neuroimaging study often involves the combination of different pipelines or the chaining of a pipeline to another one. This is the case for multimodal studies where processed outputs from a modality will be inputs for another pipeline, but it is also true for studies involving a single modality: features extracted from one or several pipelines are usually connected to statistical or machine learning frameworks. Finally, a structured organization for processed data will ease the access and sharing of data, thus improving the reproducibility of neuroimaging studies.

The CAPS format defines a hierarchy for the Clinica processed data. The idea is to include in a single folder all the results generated by the different pipelines and to organize the data following the main patterns of the BIDS specification. CAPS folders are kept separate from the raw data. Indeed, when processing data, it is very common to have the raw dataset located on a separated storage or read-only storage, while ongoing processed data are located on a separate location or on a faster data storage.

Another notion we often meet in neuroimaging studies is the notion of group, e.g. template creation from a set of subjects or statistical analysis of a population. To handle these situations, we simply add a level to the CAPS folder hierarchy. While pipeline outputs for individuals are stored in the `subjects` folder, results of group studies are stored in the `groups` folder together with the set of participants involved. For instance, an `AD` group label could be used when a template is created for a group of Alzheimer's disease patients. Any time this `AD` template is used, the `group_label` is provided to identify the pipeline outputs obtained for this group. The group `HCvsAD` could be used as `group_label` for a statistical group comparison between healthy controls (`HC`) and Alzheimer's disease patients. An illustration showing the chaining of pipelines and the creation of a group label can be found in section 6.

## 3.4    Clinica command line arguments

For each pipeline, the command line interface will require a set of arguments which can be compulsory or optional. The number of mandatory arguments is kept as small as possible, to ease its use. This set of arguments is gathered into four categories.

First, the user will be asked to provide the Clinica mandatory arguments. These arguments are in general the BIDS directory, the CAPS directory and/or the Group label, which were described in the Clinica environment section (sections 3.2 and 3.3).

Then, several options are common to every pipeline: the Clinica standard options. For instance, we can run a pipeline on a subset of participants and sessions by specifying a TSV file. Moreover, it is possible to specify the number of cores of your machine used to run pipelines in parallel thanks to the Nipype engine (Gorgolewski et al., 2011). A working directory can be specified for each pipeline. This directory gathers all the inputs and outputs of the different steps of the pipeline, which is very useful for debugging. It is especially useful in case a pipeline execution crashes to relaunch it with the exact same parameters, allowing the execution to continue from the last successfully executed node.

Other parameters, specific to each pipeline, are gathered in the category "Optional parameters". For instance, when applying a smoothing filter with a specific full width at half maximum (FWHM), this parameter can be specified.

Finally, advanced parameters for users with good knowledge of the pipeline itself or of the software behind the pipeline will be gathered in the category "Advanced pipeline options".

## 3.5    List of atlases available in Clinica

Depending on the modality studied and the type of analysis (voxel-based or surface-based), different atlases can be used to generate regional features. These atlases are briefly listed below, and the reader can refer to the documentation available on the website for further details.

When performing volumetric processing of T1w MRI and PET images, as done in the `t1-volume*` and `pet-volume` pipelines, atlases defined in MNI space containing regions covering the whole cortex and the main subcortical structures available are used (Samper-González et al., 2018), currently AAL2 (Tzourio-Mazoyer et al., 2002), AICHA (Joliot et al., 2015), Hammers (Hammers et al., 2003; Gousias et al., 2008), LPBA40 (Shattuck et al., 2008), and Neuromorphometrics[14].

When running the `dwi-dti` pipeline, the JHUDTI81 (Wakana et al., 2007; Hua et al., 2008) and JHUTracts[0|25|50] (Mori et al., 2005) atlases[15], included in FSL (Jenkinson et al., 2012), defined

---

[14] www.neuromorphometrics.com

[15] https://fsl.fmrib.ox.ac.uk/fsl/fslwiki/Atlases

in MNI space, are used. JHUDTI81 contains 48 white matter tract labels and JHUTracts[0|25|50] contains 20 white matter probabilistic tract labels with a 0%, 25% and 50% threshold.

Moreover, surface atlases are used when processing T1w MRI (respectively PET images) with the `t1-freesurfer*` (respectively `pet-surface*`) pipelines. Currently, Clinica provides the Desikan-Killiany (Desikan et al., 2006) atlas, which divides the cerebral cortex into gyri and contains 34 regions per hemisphere, and the Destrieux (Destrieux et al., 2010) atlas, which divides the cerebral cortex into gyri and sulci and contains 74 regions per hemisphere.

### 3.6    Continuous Integration, testing and package distribution

The source code of the Clinica's platform follows the most standard current practices for software development. The code is hosted in a publicly available platform[16] and it uses a version control system. A rigorous code review is performed for every contribution. The project has adopted a commonly used workflow for development and the code is tested at different stages under controlled conditions. In order to do this, several pipelines are executed by the continuous integration setup at different levels.

For each contribution proposal:

- The most recent commit pushed to the repository triggers a first iteration of the test suite. This first round validates the package environment, the installation process and the correct instantiation of the main tools proposed by Clinica.
- A draft of the documentation is written and published once the first iteration is over.

Then, the contribution proposal is reviewed and validated by a peer. Subsequently:

- Nightly tests ensure that new contributions do not introduce regressions in the results of the software. This second iteration of the test suite runs the full set of Clinica's functionalities and, due to the long processing time, they are executed once a day.
- Package construction and deployment is automatized by adding a tag with the version number to the VCS. Versioned packages are published in the Python Package Index[17].

The management of the continuous integration system is handled by a master server that creates the link between the code repository and the different virtual machines that execute the continuous integration tasks. Virtual machines are configured with Linux and macOS operating systems.

Outputs from the continuous integration process are publicly available and contributors can easily consult them. Due to legal restrictions, the datasets used during the continuous integration cannot be publicly distributed but detailed instructions on how to obtain them are provided on demand.

---

[16] https://github.com/aramis-lab/clinica/

[17] https://pypi.org/project/clinica/

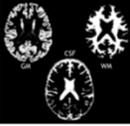
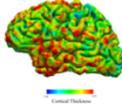
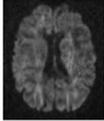
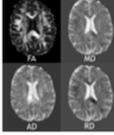
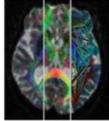
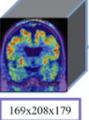
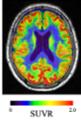
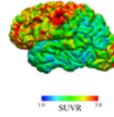

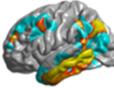
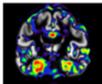
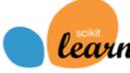
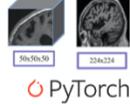
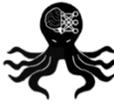

**Anatomical MRI**

**t1-linear**
Bias field correction, affine registration and cropping
Dependencies: ANTs

• T1 MRI on ICBM 2009c nonlinear symmetric template
• Used as input for deeplearning-prepare-data

169x208x179

**t1-volume**
Tissue segmentation (GM, WM, CSF), inter-subject registration using Dartel, spatial normalization to standard space (MNI)
Dependencies: SPM, CAT12

• Voxel-based features (GM, WM, CSF)
• Regional features (average GM) using atlases (currently AAL2, AICHA, Hammers, LPBA40, Neuromorphometrics)

**t1-freesurfer**
**t1-freesurfer-longitudinal**
Cortical surface extraction, segmentation of subcortical structures, cortical thickness estimation, spatial normalization to standard space (FsAverage)
Dependencies: FreeSurfer

• Surface-based features (cortical thickness)
• Regional features (average cortical thickness) using atlases (currently Desikan, Destrieux)

**Diffusion MRI**

**dwi-preprocessing**
Correction of raw DWI data
Dependencies: FSL, ANTs, Convert3D

• EPI correction using phase-difference map fieldmap or T1w ("fieldmap-less")
• Prerequisite for dwi-dti and dwi-connectome pipelines

**dwi-dti**
Extraction of DTI-based measures, normalization to standard space (MNI)
Dependencies: FSL, ANTs, MRtrix3

• Voxel-based features (FA, MD, AD, RD)
• Regional features (average FA, MD, AD, RD) using atlases (JHUDTI81, JHUTracts)

**dwi-connectome**
Tractography & connectome
Dependencies: FreeSurfer, FSL, MRtrix

• Probabilistic tractography
• Structural connectome using atlases (currently Desikan, Destrieux)

**PET**

**pet-linear**
Affine registration, intensity normalization and cropping
Dependencies: ANTs

• PET on ICBM 2009c nonlinear symmetric template
• Used as input for deeplearning-prepare-data

169x208x179

**pet-volume**
Registration to T1 MRI, partial volume correction, spatial normalization to standard space (MNI), intensity normalization
Dependencies: SPM, PETPVC, CAT12

• Voxel-based features (e.g. FDG uptake, amyloid uptake)
• Regional features (average FDG, amyloid uptake) using atlases (currently AAL2, AICHA, Hammers, LPBA40, Neuromorphometrics)

SUVR

**pet-surface**
**pet-surface-longitudinal**
Registration to T1 MRI, intensity normalization, partial volume correction, projection to cortical surface, spatial normalization to standard space (FsAverage)
Dependencies: FreeSurfer, FSL, SPM, PETPVC

• Surface-based features (e.g. FDG uptake, amyloid uptake)
• Regional features (average cortical thickness) using atlases (currently Desikan, Destrieux)

SUVR

**Statistics**

**statistics-volume**
Voxel-based mass-univariate analysis with SPM
Dependencies: SPM, Matlab

• Voxel-based features from t1-volume or pet-volume pipelines
• Group comparison using GLM

**statistics-surface**
Surface-based mass-univariate analysis with SurfStat
Dependencies: Matlab

• Surface-based features from t1-freesurfer or pet-surface pipelines
• Group comparison or correlations analysis using GLM

**Machine Learning**

**machinelearning-prepare-spatial-svm**
Preparation of T1 MRI and PET data for spatially regularized SVM
Dependencies: None

• Regularization that accounts for the spatial and anatomical structure of neuroimaging data leading to a more regular and anatomically interpretable decision function.
• Used as input for machine learning classification

**(No command line interface)**
Classification based on machine learning
Dependencies: None

• Voxel-based, surface-based or regional features
• Classifications (SVM, ℓ2 logistic regression, random forest) using cross-validations (K-fold, repeated K-fold, repeated hold-out)

learn

**Deep Learning**

**deeplearning-prepare-data**
Convert features extracted by Clinica to PyTorch tensors
Dependencies: None

• 3D images, 3D patches or 2D slices from t1-linear or t1-volume pipelines
• Tensors for PyTorch

PyTorch

**clinicadl**
Reproducible classification of Alzheimer's disease using deep learning
Installation: pip install clinicadl

• Tensors from deeplearning-prepare-data pipeline
• CNN Classification using K-fold cross-validation or single data split
• Features such as random search, interpretability and meta-data management

**Figure 2** List of the pipelines currently available in Clinica with their dependencies and outputs. Explanations regarding the atlases can be found in section 3.5. GM: gray matter; CSF: cerebrospinal fluid; WM: white matter; FA: fractional anisotropy; MD: mean diffusivity; AD: axial diffusivity; RD: radial diffusivity, SVM: Support Vector Machine; ICBM: International Consortium for Brain Mapping.

## 4 Image processing pipelines (clinica run)

This section gives a brief description of the different pipelines currently provided by Clinica as well as the types of features Clinica can produce. An illustrative summary of the pipelines can be found in Figure 2.

For technical details, we refer the reader to the online user documentation available on the Clinica website where a longer description of each pipeline is provided.

### 4.1 Anatomical MRI

#### 4.1.1 Linear processing of T1-weighted MR images (t1-linear)

The `t1-linear` pipeline performs a set of steps in order to affinely align T1w MR images to the MNI space using the ANTs software package (Avants et al., 2014a, 201). These steps include: bias field correction using N4ITK (Tustison et al., 2010); affine registration to the MNI152NLin2009cSym template (Fonov et al., 2009, 2011) in MNI space with the SyN algorithm (Avants et al., 2008a); cropping of the registered images to remove the background.

This pipeline was designed to be a prerequisite for the `deeplearning-prepare-data` pipeline.

#### 4.1.2 Processing of T1-weighted MR images for volume analyses using SPM (t1-volume*)

The `t1-volume*` pipelines extract voxel-based anatomical features from T1w MR images. Specifically, they perform segmentation of tissues (gray matter (GM), white matter (WM), cerebrospinal fluid (CSF)), normalization to MNI space and computation of regional measures using atlases. Their main outputs are voxel-based maps of tissue density and average measures within cortical regions stored as TSV files.

To that purpose, the pipeline wraps the Segmentation, Run Dartel and Normalise to MNI Space routines implemented in SPM (Ashburner, 2012). First, the Unified Segmentation procedure (Ashburner and Friston, 2005) is used to simultaneously perform tissue segmentation, bias field correction and spatial normalization of the input image. Next, a group template is created using DARTEL, an algorithm for diffeomorphic image registration (Ashburner, 2007), from the subjects' tissue probability maps on the native space, usually GM, WM and CSF, obtained at the previous step. The DARTEL to MNI method (Ashburner, 2007) is then applied, providing a registration of the native space images into the MNI space.

#### 4.1.3 Processing of T1-weighted MR images for surface analyses using FreeSurfer (t1-freesurfer; t1-freesurfer-longitudinal)

The `t1-freesurfer` pipeline is mainly a wrapper of the `recon-all` tool of FreeSurfer (Fischl, 2012). It performs segmentation of subcortical structures (Fischl et al., 2002, 2004a), extraction of cortical surfaces, cortical thickness estimation (Fischl and Dale, 2000), spatial normalization onto the FreeSurfer surface template (FsAverage) (Fischl et al., 1999), and parcellation of cortical regions using the Desikan and Destrieux atlases (Fischl et al., 2004b). Its main outputs are surface-based cortical thickness features and regional statistics (e.g. regional volume, mean cortical thickness).

The `t1-freesurfer-longitudinal` pipeline processes a series of images acquired at different time points for the same subject with the longitudinal FreeSurfer stream (Reuter et al., 2012) to increase the accuracy of volume and thickness estimates. It does so in a single command consisting of two consecutive steps: 1) within-subject template creation (`recon-all -base` command) to produce an unbiased template image from the different time points using robust and inverse consistent registration (Reuter et al., 2010); 2) longitudinal correction (`recon-all -long` command): segmentation, surface extraction and computation of measurements at each time point.

## 4.2 Diffusion MRI

### 4.2.1 DWI pre-processing (dwi-preprocessing-*)

The `dwi-preprocessing-*` pipelines correct diffusion-weighted MRI (DWI) datasets for motion, eddy current, magnetic susceptibility and bias field distortions, assuming that the data have been acquired using an echo-planar imaging (EPI) sequence.

Due to the heterogeneity in acquisitions of fieldmaps and techniques to correct magnetic susceptibility distortions, several pipelines are proposed. Currently, Clinica can handle DWI datasets with fieldmap data containing a phase-difference map (case 'phase-difference map and at least one magnitude image' in the BIDS specifications[18]) (`dwi-preprocessing-using-fmap`) and DWI datasets with no extra data (`dwi-preprocessing-using-t1`), which is the case of the public Alzheimer's Disease Neuroimaging Initiative (ADNI)[19] dataset for instance.

In all cases, motion and eddy motion corrections are performed with the FSL software (Jenkinson et al., 2012) using the `eddy` tool (Andersson and Sotiropoulos, 2016) with the replace outliers (`--repol`) option (Andersson et al., 2016) while bias field is corrected with the ANTs N4 bias correction (Tustison et al., 2010). Regarding susceptibility correction, the FSL prelude/fugue tools were used for the `dwi-preprocessing-using-fmap` pipeline and the ANTs SyN registration algorithm (Leow et al., 2007; Avants et al., 2008a) for the `dwi-preprocessing-using-t1` pipeline. The outputs of the pipelines are the corrected DWI datasets and a brain mask of the b=0 image.

These pipelines are prerequisites for the `dwi-dti` and `dwi-connectome` pipelines.

### 4.2.2 Computation of DTI, DTI-scalar maps and ROI analysis (dwi-dti)

The `dwi-dti` pipeline extracts voxel-based features from diffusion tensor imaging (DTI), namely the fractional anisotropy (FA), mean diffusivity (MD), axial diffusivity (AD) and radial diffusivity (RD) using MRtrix3 (Tournier et al., 2012). Then, the DTI-derived scalar maps (FA, MD, AD, RD) are normalized with ANTs (Avants et al., 2008a) onto an FA-atlas with labelled tracts. Its main outputs are voxel-based maps from DTI and average measures within tracts stored as TSV files.

---

[18] http://bids.neuroimaging.io/

[19] http://adni.loni.usc.edu/

### 4.2.3 Computation of fiber orientation distributions, tractogram and structural connectome (dwi-connectome)

The `dwi-connectome` pipeline computes a weighted graph encoding anatomical connections between a set of brain regions from corrected DWI datasets. To that aim, it relies on the MRtrix3 (Tournier et al., 2012) software to compute the constrained spherical deconvolution diffusion model (Tournier et al., 2007), perform probabilistic tractography (Tournier et al., 2010) and computes a connectome using the Desikan & Destrieux atlases from FreeSurfer (Fischl, 2012). Its main outputs are the diffusion model, the whole-brain tractography and the connectivity matrices.

## 4.3 Positron emission tomography

Currently, Clinica is supporting amyloid and FDG PET data but other tracers will be added in the future.

### 4.3.1 Linear processing of PET images (pet-linear)

The `pet-linear` pipeline performs a spatial normalization to the MNI space and intensity normalization of PET images. The first step of the pipeline is an affine registration to the MNI152NLin2009cSym template (Fonov et al., 2009, 2011) in MNI space with the SyN algorithm (Avants et al., 2008b) from the ANTs software package (Avants et al., 2014b). Then, the registered image intensity is normalized using the mean intensity in reference regions resulting in a standardized uptake value ratio (SUVR) map. The normalized imaged is finally cropped to remove the background.

### 4.3.2 Processing of PET images for volume analyses (pet-volume)

The `pet-volume` pipeline extracts voxel-based features from PET data. Specifically, it performs intra-subject registration of the PET image into the space of the subject's T1w MR image using SPM (Ashburner, 2012). Optionally, partial volume correction (PVC) can be applied thanks to the PETPVC toolbox (Thomas et al., 2016). Then, inter-subject spatial normalization of the PET image into MNI space is performed based on the DARTEL deformation model of SPM (Ashburner, 2007) and intensity normalization is done using the average PET uptake in a reference region resulting in a SUVR map. Its main outputs are voxel-based maps of SUVR and average measures within cortical regions.

### 4.3.3 Processing of PET images for surface analyses (pet-surface; pet-surface-longitudinal)

The `pet-surface` pipeline extracts the PET signal and projects it onto the cortical surface using the approach described in (Marcoux et al., 2018). More precisely, it performs co-registration of PET and T1w MRI, intensity normalization, PVC with the PETPVC toolbox (Thomas et al., 2016), robust projection of the PET signal onto the subject's cortical surface, parcellation of the cortical regions using the Desikan and Destrieux atlases and spatial normalization onto the FreeSurfer (Fischl, 2012) surface template (FsAverage). Its main outputs are surface-based PET uptake and regional statistics (mean PET uptake) stored as TSV files.

The `pet-surface-longitudinal` pipeline performs the same steps as the `pet-surface` pipeline except that the cortical and white surfaces are estimated with the longitudinal pipeline of FreeSurfer (Reuter et al., 2012).

## 4.4 Statistics

### 4.4.1 Voxel-based mass-univariate analysis with SPM (statistics-volume)

The `statistics-volume` pipeline performs statistical analysis on volume-based features using the general linear model (GLM) and random field theory (Worsley et al., 2009). To that aim, the pipeline wraps the statistical analysis toolbox implemented in SPM. Volume-based measurements can be gray matter maps from the `t1-volume` pipeline or PET measurements from the `pet-volume` pipeline. Currently, statistical analysis includes only group comparison. The pipeline is divided into two subpipelines: `statistics-volume` outputs an SPM-based report with uncorrected T-statistic maps as well as the computed thresholds for family-wise error or false discovery rate correction at the voxel or the vertex level; `statistics-volume-correction` outputs corrected T-statistic maps with each of the aforementioned corrections.

### 4.4.2 Surface-based mass-univariate analysis with SurfStat (statistics-surface)

The `statistics-surface` pipeline performs statistical analysis on surface-based features using the GLM. To that aim, the pipeline relies on the Matlab toolbox SurfStat designed for statistical analyses of univariate and multivariate surface and volumetric data using the GLM (Worsley et al., 2009). Surface-based measurements are analyzed on the FsAverage surface template from FreeSurfer. The pipeline can handle cortical thickness from the `t1-freesurfer` pipeline or PET measurements from the `pet-surface` pipeline. Currently, statistical analysis includes group comparison and correlation. The main outputs are p-value maps with different corrections for multiple comparisons (at the cluster or the vertex level using random field theory, or using the false discovery rate) and a JSON file summarizing the specified model.

## 4.5 Machine Learning

### 4.5.1 Classification based on machine learning (no command line)

Clinica provides a modular way to perform classification based on machine learning. To build their own classification pipeline, the user can combine three modules based on the scikit-learn library (Pedregosa et al., 2011):

- Input module. The user can select the inputs from the features available in the CAPS directory, such as gray matter maps obtained from T1w MR images, or SUVR maps obtained from FDG PET images.
- Algorithm module. The user can choose between different classifiers, currently support vector machine (SVM), logistic regression and random forest.

- Validation module. Several cross-validation (CV) methods are available: k-fold CV, repeated k-fold CV and repeated hold-out CV.

Note that no command line interface is available for these specific tools. They need to be used within Python code, for instance within a notebook (an example of such notebook is provided at: https://github.com/aramis-lab/AD-ML/blob/master/Generic_Version/Experiments.ipynb).

The outputs are: the estimated model parameters, the optimal value of the hyper-parameters (if any), a set of metrics regarding the classification performance (balanced accuracy, AUC, accuracy, sensitivity, specificity…) as well as the predicted class for each subject thereby allowing to calculate any additional performance metric.

More details regarding the different modules and a description of the way they can be used to perform reproducible evaluation of classification methods in Alzheimer's disease can be found in (Samper-González et al., 2018) and its dedicated repository[20].

### 4.5.2 Spatially-regularized support vector machine (machinelearning-prepare-spatial-svm)

The `machinelearning-prepare-spatial-svm` pipeline allows the preparation of T1w MRI and PET data to perform classification with an SVM with spatial and anatomical regularization (Cuingnet et al., 2013). In this approach, the standard regularization of the SVM is replaced with a regularization that accounts for the spatial and anatomical structure of neuroimaging data. More specifically, it is regularized with respect to the tissue maps (GM, WM, CSF). As a result, the decision function learned by the algorithm will be more regular and anatomically interpretable. Because the SVM is a kernel method, the spatial/anatomical regularization is done as a pre-processing on the feature maps and the result can then be fed to a standard linear SVM. This pipeline outputs spatially regularized maps that can then be entered into a standard SVM, providing the same type of outputs as in section 4.5.1.

### 4.6 Deep Learning

### 4.6.1 Prepare data for deep learning (deeplearning-prepare-data)

The `deeplearning-prepare-data` pipeline allows the preparation of data for subsequent training or inference of deep learning models. To that aim, it uses the outputs from `t1-linear` or `pet-linear` pipelines. Specifically, 3D images, 3D patches or 2D slices can be extracted and converted into PyTorch tensors (Paszke et al., 2019). The outputs are thus the corresponding extracted images, patches or slices as .pt (PyTorch tensors) files.

---

[20] https://github.com/aramis-lab/AD-ML

### 4.6.2 Training and validation of deep learning models (clinicadl)

The training and validation of deep learning models based on Clinica outputs can be performed using a dedicated Python library: ClinicaDL[21]. This extension of Clinica contains essential features for deep learning application to 3D medical images:

- modules to split data avoiding data leakage, which is a major problem in the domain (Wen et al., 2020);
- a training method for autoencoders, CNN and a multi-CNN framework which allows the use of other networks trained with ClinicaDL for transfer learning;
- a testing function to evaluate the performance of classifiers on independent test sets;
- saliency maps generation (Simonyan et al., 2013) extensively used to interpret the outputs of deep learning networks;
- basic network architecture search tools, such as random search utilities and methods to generate trivial synthetic datasets for architecture debugging.

This library is documented in an independent documentation[22]. An online tutorial[23] allows beginners to better understand these functionalities by testing them locally, or on Google Colab if they do not have access to sufficient computational resources. Each function of ClinicaDL has specific outputs (training functions output models, classification functions output performance metrics and classification results…). We refer the reader to this specific documentation for an exhaustive description.

## 5    Clinica utilities

## 5.1    Conversion of neuroimaging datasets into a BIDS hierarchy (clinica convert)

Clinica provides tools to curate several publicly available neuroimaging datasets and automatically convert them into the BIDS standardized data structure. This section explains what the user needs to download prior to running the converter and the rationale behind the selection of data when multiple acquisitions or pre-processing steps are available. For all converters, the user only needs to download the dataset. All subsequent conversion steps are performed automatically (no user intervention is required) and use parallelization for faster processing. For further details, the reader can refer to (Samper-González et al., 2018). Clinica currently provides converters for the following studies: ADNI, AIBL, NIFD and OASIS. We plan to continuously add new converters for other studies. In addition, Clinica can of course be used with any other dataset, provided that it has previously been converted to BIDS by the user.

---

[21] https://github.com/aramis-lab/AD-DL

[22] https://clinicadl.readthedocs.io/en/latest/

[23] https://aramislab.paris.inria.fr/clinicadl/tuto/intro.html

### 5.1.1 Conversion of the ADNI dataset to BIDS (adni-to-bids)

The ADNI to BIDS converter requires the user to have downloaded all the ADNI study data (tabular data in CSV format) and the imaging data of interest. Note that the downloaded files must be kept exactly as they were downloaded. The imaging modalities currently being converted to BIDS include T1w MRI, FLAIR, DWI, fMRI, FDG PET, PiB PET, Florbetapir (AV45) PET and Flortaucipir (AV1451) PET. Clinical data are also converted to BIDS. They include data that do not change over time, such as the subject's sex, education level or diagnosis at baseline, as well as session-dependent data, such as the clinical scores. The clinical data being converted are defined in a spreadsheet that is available with the code of the converter. The user can easily modify this file if they want to convert additional clinical data.

### 5.1.2 Conversion of the AIBL dataset to BIDS (aibl-to-bids)

As for ADNI, the AIBL to BIDS converter requires the user to have downloaded the AIBL non-imaging data (tabular data in CSV format) and the imaging data of interest. For each AIBL participant, the T1w MRI and the Florbetapir, PiB and Flutemetamol PET images are converted. As for the ADNI converter, clinical data converted to BIDS are defined in a spreadsheet available with the code of the converter, which the user can modify.

### 5.1.3 Conversion of the NIFD dataset to BIDS (nifd-to-bids)

As for ADNI, the NIFD to BIDS converter requires the user to have downloaded the NIFD imaging data alongside the corresponding clinical data in CSV format. For each NIFD participant, the T1w MRI, FLAIR, PiB PET and FDG PET images are converted. The clinical data conversion is as described in the previous sections.

### 5.1.4 Conversion of the OASIS dataset to BIDS (oasis-to-bids)

As for ADNI, the OASIS to BIDS converter requires the user to have downloaded the OASIS-1 imaging data and the associated CSV file. For each subject, among the multiple T1w MR images available, we select the average of the motion-corrected co-registered individual images resampled to 1 mm isotropic voxels. The clinical data are converted as described in the previous sections.

### 5.1.5 Syntax to run the converters

After having downloaded the clinical and imaging data of one of these studies, the conversion of a dataset into BIDS is performed using the following syntax:

```
clinica convert <dataset>-to-bids dataset_directory clinical_data_directory
bids_directory
```

where `<dataset>-to-bids` can be `adni-to-bids`, `aibl-to-bids`, `nifd-to-bids` or `oasis-to-bids`.

## 5.2    Data handling tools (clinica iotools)

We also propose a set of tools that allows the user to handle BIDS and CAPS datasets. For the moment, there are five different commands:

- `center-nifti`: This command takes a BIDS directory as input and outputs the same BIDS with centered NifTI: the NifTI headers are modified to set the origin of the coordinate system at the center of the image. This correction is crucial for SPM which is especially sensitive to NifTI files whose origin does not correspond to the center of the image.
- `check-missing-modalities`: This command checks missing modalities in a BIDS directory.
- `check-missing-processing`: This command checks the outputs in a CAPS directory.
- `create-subjects-visits`: This command generates a list of subjects with their sessions based on a BIDS directory and stores the outputs in a TSV file.
- `merge-tsv`: This command merges all the tabular data including the clinical data of a BIDS directory and the regional features from a CAPS directory (e.g. mean GM density in AAL2 atlas) into a single TSV file. This file can then be easily plugged into machine learning tools via Clinica or other statistical/machine learning software packages.

# 6   Usage example

In this section, we propose to show how Clinica can be used to perform a group comparison of FDG PET data projected on the cortical surface between patients with Alzheimer's disease and healthy controls from the ADNI database. An illustrative summary of this example can be found in Figure 3.

To download the ADNI dataset, it is necessary to register to the LONI Image & Data Archive[24], a secure research data repository, and request access to the ADNI dataset through the submission of an online application form. Both the imaging and clinical data need to be downloaded, each to a folder that we will call `imaging_data_dir` and `clinical_data_dir`, respectively. The following command can be used to convert the T1 and FDG PET data of the ADNI dataset into BIDS:

```
clinica convert adni-to-bids imaging_data_dir clinical_data_dir ADNI_BIDS
--modalities T1 PET_FDG
```

where the `ADNI_BIDS` folder contains the conversion of ADNI into BIDS. We can now start processing the data. First, we need to extract the cortical surfaces from each anatomical image. To do so, we simply need to type on the terminal the following command:

```
clinica run t1-freesurfer ADNI_BIDS ADNI_CAPS
```

where the output data will be stored in the `ADNI_CAPS` folder. After visual inspection of the generated outputs, the FDG PET data can be projected onto the cortex. The command line will be:

```
clinica run pet-surface ADNI_BIDS ADNI_CAPS fdg pons pvc_psf.tsv
```

where `fdg` is the label given to the PET acquisition, `pons` is the reference region for the SUVR map computation and `pvc_psf.tsv` is the TSV file containing PSF information for each PET image. Finally, we can perform group comparison of cortical FDG PET data after having checked the outputs. The demographic information of the population studied will be stored in a TSV file, looking as follows:

```
participant_id      session_id      group      age      sex

sub-ADNI094S2201    ses-M00         HC         63.7     Female

sub-ADNI098S4018    ses-M00         HC         76.1     Male

sub-ADNI023S4020    ses-M00         HC         66.5     Male

sub-ADNI031S4021    ses-M00         HC         66.5     Male

sub-ADNI094S1397    ses-M00         AD         55.1     Female

sub-ADNI094S1402    ses-M00         AD         69.3     Male

sub-ADNI128S1409    ses-M00         AD         65.9     Male

sub-ADNI128S1430    ses-M00         AD         83.4     Female

...
```

where participants with Alzheimer's disease (respectively healthy controls) have the `AD` label (respectively `HC` label) in the `group` column. We will call this file `ADvsHC_participants.tsv`. Using age and sex as covariates, the command line will be:

---

```
clinica run statistics-surface ADNI_CAPS ADvsHC pet-surface \
    group_comparison ADvsHC_participants.tsv group --covariates "age + sex"
```

The results of the statistical analysis will be stored in the `ADNI_CAPS/groups/group-ADvsHC` folder.

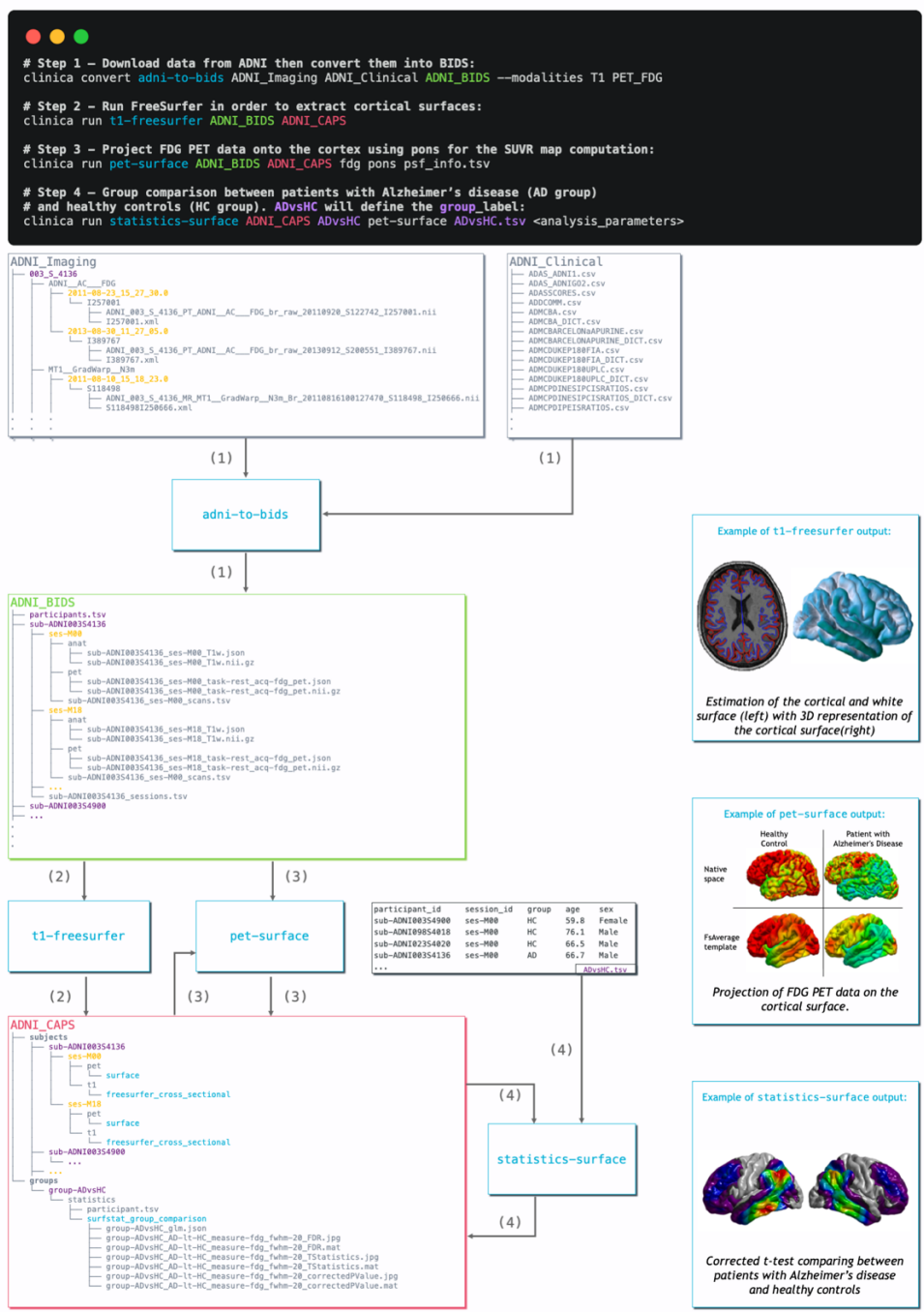

**Figure 3** Diagram illustrating the Clinica pipelines involved when performing a group comparison of FDG PET data projected on the cortical surface between patients with Alzheimer's disease and healthy controls from the ADNI database. First, clinical and neuroimaging data are downloaded from the ADNI website and data are converted into BIDS with the `adni-to-bids` tool from Clinica (1). Estimation of the cortical and white surface is then produced by the `t1-freesurfer` pipeline in a single command line (2). Afterwards, FDG PET data can be projected on the subject's cortical surface and normalized to the FsAverage template from FreeSurfer using the `pet-surface` pipeline (3). Finally, a TSV file with demographic information of the population studied is given to the `statistics-surface` pipeline to generate the results of the group comparison between patients with Alzheimer's disease and healthy controls (4).

# 7 Discussion

We proposed a software platform that aims at making clinical neuroscience easier and more reproducible. Clinica automates the processing of pipelines involving several neuroimaging modalities (currently, anatomical MRI, diffusion MRI, and PET) as well as statistics, machine learning and deep learning tools. Additionally, Clinica provides tools to convert public neuroimaging datasets focused on dementia (ADNI, AIBL, OASIS and NIFD) into the BIDS standard, and tools to handle raw (BIDS) and processed datasets. The use of the BIDS standard as the only prerequisite on the data and the unified command line interface across the pipelines ease the processing automation. The image analysis automation is also improved by the use of the CAPS hierarchy, which facilitates the chaining of pipelines. The main target audience of Clinica is neuroscientists or clinicians conducting clinical neuroscience studies involving multimodal imaging, and researchers developing advanced machine or deep learning algorithms.

The last three decades witnessed the development of many software packages for the processing of neuroimaging data. A first category of packages comprises those implementing innovative image processing methodologies (e.g. tissue segmentation, registration). Many tools fall into this category, for instance SPM (Friston et al., 2007), AFNI (Cox, 1996), FreeSurfer (Fischl, 2012), FSL (Jenkinson et al., 2012), PETPVC (Thomas et al., 2016), Camino (Cook et al., 2005), Dipy (Garyfallidis et al., 2014), DTI-TK[25], MRtrix (Tournier et al., 2012) or ANTs (Avants et al., 2014a). Some of these tools cover a variety of modalities while others focus on a specific one (diffusion MRI for Camino, Dipy, DTI-TK and MRtrix; fMRI for AFNI; PET for PETPVC). However, performing a multimodal study can be difficult because one needs to combine tools from different packages. This results in complex pipelines which can be difficult to build, maintain and distribute. Even when analyzing a single modality, one often wants to combine tools from different packages, thus facing similar difficulties. Combination of tools is made even more difficult by the fact that the input and output data are organized differently by each tool.

Efforts of the community have alleviated several of these difficulties. The NeuroDebian community[26] (Halchenko and Hanke, 2012) aims to provide and ease the installation of a large collection of software packages for the Debian distribution. The Nipype (Gorgolewski et al., 2011) system facilitates the building of complex pipelines through the wrapping of tools in Python. BIDS (Gorgolewski et al., 2017) provides a standard for organizing data. BIDS-Apps provides versions of software packages using BIDS for data organization. More generally, the NIPY community aims to provide a comprehensive set of tools for the analysis of neuroimaging data in a single language, Python. However, many useful tools and packages remain outside of the NIPY scope, being written in different languages. As mentioned above, Nipype allows wrapping these heterogeneous tools. It is a powerful and particularly useful tool for that aim. However, the building of pipelines remains left to the user. This requires substantial development efforts. Clinicians/neuroscientists often do not have the necessary programming expertise, while researchers in machine learning often do not have the necessary neuroimaging expertise. Therefore, Nipype and Clinica do not have the same objectives. Nipype provides a powerful way to write pipelines. As such, it is particularly flexible but requires some programming expertise. Clinica, on the other hand, offers a set of predefined pipelines: it is thus easier to use but less flexible.

---

[25] http://dti-tk.sourceforge.net

[26] http://neuro.debian.net/

There are also software packages that integrate different tools within a single environment. This is for example the case of BCBtoolkit (Foulon et al., 2018), BrainVISA (Cointepas et al., 2001), BrainSuite[27], BrainLife[28] (Avesani et al., 2019), Flywheel[29], fMRIPrep (Esteban et al., 2019), MIBCA[30] (Ribeiro et al., 2015) or Pypes (Savio et al., 2017). Clinica falls within this category. It shares some characteristics with these tools but also has important differences. The BCBtoolkit wraps neuroimaging software packages from the community but also highlights new methodological developments to evaluate brain disconnections. BCBoolkit does not use any pipelining system but instead wraps bash scripts that are then made available through a GUI. The BrainVISA platform, even though it also wraps some existing tools, mainly provides innovative tools for the analysis of human or animal brain imaging data. Moreover, it includes its own pipelining system while Clinica relies on the community effort Nipype. Brainlife is a comprehensive and user-friendly cloud-based platform for neuroscience analysis and provides an interface for composable integration of different neuro-imaging tools. Unlike Clinica however, Brainlife is not available as a Python package. Additionally, both the upstream processing provided by Clinica (curation and transformation of common research dataset into BIDS) and downstream (preparation for ML pipelines) are not the focus of Brainlife. BrainSuite does not wrap existing tools but provides a set of innovative tools for the analysis of neuroimaging data. It can be executed using a GUI, a command line, a Nipype interface, or as a BIDS App. However, BrainSuite is limited to the processing of MRI images only. Flywheel is a cloud-based platform to capture, curate and process medical data. In the spirit of Brainlife, it allows the composition of different components through a web-based interface. However, unlike Clinica, Flywheel is a proprietary software and does not share its source code with the community. fMRIPrep combines software components using Nipype (Gorgolewski et al., 2011) to provide a robust pipeline for pre-processing of fMRI data. It assumes input data to follow the BIDS standard (Gorgolewski et al., 2016) and outputs are organized following the ongoing BIDS-derivatives initiative. As the name implies, fMRIPrep focuses on fMRI data. MIBCA is a toolbox focused on brain connectivity analysis using multimodal imaging. It is developed in MATLAB and provides pipelines for popular neuroimaging software. Pypes is probably the closest in spirit to Clinica: it also focuses on the integration of existing tools into a set of reusable pipelines built with Nipype. The user needs to specify a configuration file to describe the input data even for BIDS datasets. The output data will then follow the same structure as the input data and can be chained to other pipelines from Pypes. Note that the ability to chain pipelines exists for a limited number of pipelines in the BIDS-App version of BrainSuite and is likely to be more present in BIDS Apps with the advent of BIDS-derivatives. Although Clinica does not provide a GUI, efforts were made to simplify as much as possible its command line interface, which is feasible thanks to the autocompletion, the structured organization of data and the documentation which was designed in order to be readable by a newcomer.

Machine learning is now widely used in the neuroimaging community, for cognitive neuroscience or computer-aided diagnosis applications. However, applying such approaches to neuroimaging data can be difficult for newcomers. Conversely, researchers in machine learning are often interested in applying and validating their approaches to clinical neuroimaging problems (e.g. diagnosis, prognosis). However, they often lack the necessary knowledge for preprocessing neuroimaging data and extracting features. Nilearn (Abraham et al., 2014) has allowed major progress in that direction by providing in a single environment tools for preprocessing, data manipulation, feature extraction and machine learning wrapping the scikit-learn library (Pedregosa et al., 2011). However, Nilearn currently mostly targets cognitive neuroimaging and mainly deals with functional neuroimaging. On the other hand, Clinica is dedicated to clinical neuroimaging studies, such as biomarker design and clinical decision support systems. As a result, Clinica does not aim to deal with task-based fMRI and is currently mostly focused on the analysis of T1w MRI, diffusion MRI and PET data. Deep learning methods are also increasingly applied to neuroimaging. Even though such methods usually do not rely on pre-extracted features, they will still require some preprocessing and data conversion tools in order to use neuroimaging data in frameworks such as PyTorch. Clinica provides such tools.

---

[27] http://brainsuite.org

[28] https://brainlife.io/

[29] https://flywheel.io/product/

[30] http://www.mibca.com/

Reproducibility has been highlighted as a major challenge in many scientific fields including neuroimaging (Poldrack et al., 2017). This problem has also been highlighted in machine learning for healthcare in general (McDermott et al., 2019) and for brain diseases in particular (Samper-González et al., 2018; Wen et al., 2021). Clinica aims to make reproducible research easier to perform. To that purpose, it combines: 1) use of a community standard for inputs; 2) the definition of a standardized organization for outputs; 3) standardized ways to extract features; 4) extensive software testing. As previously mentioned, it extensively relies on community achievements such as the BIDS standard and Nipype.

As mentioned above, the main target audience of Clinica is neuroscientists or clinicians and researchers developing machine algorithms applied to neuroimaging. Clinica can also be useful to neuroimaging researchers even though they often have their own customized sets of pipelines that are specifically tailored to their needs. Nevertheless, it can still be useful to them by providing a comprehensive set of pipelines for various imaging modalities. This will for instance be beneficial if they need to perform a study for a specific imaging modality for which they do not already have their own pipelines. In addition, they will benefit from a standardized file organization and easy connection with machine learning or deep learning tools. As for the clinicians, we acknowledge that the use of a command line interface as well as the need to install third-party software may make it difficult for them to use Clinica. Our clinical collaborators were nevertheless able to successfully use Clinica but this is clearly a limitation. In the future, we aim to implement a visual dashboard that would simultaneously allow executing the pipelines and performing quality control (QC) as well as to ease or avoid the installation of third-party software.

Clinica pipelines rely on third-party dependencies whose development and versioning are outside of our control. To avoid potential incompatibilities, each pipeline declares its (versioned) dependencies in the corresponding pipeline metadata, which are checked at runtime by Clinica. Nevertheless, the installation of third-party software may be difficult for the user and we acknowledge that this is a limitation. In order to limit as much as possible these difficulties, the Clinica documentation provides detailed instructions regarding the available download channels and installation steps required. Moreover, there is an on-going effort to make our third-party dependencies available through Anaconda, which will enable simpler deployments of Clinica in the near future. Nevertheless, some of the dependencies (e.g. FSL, Freesurfer) cannot be distributed in that manner due to license issues. In the future, we plan to distribute a Docker version of Clinica as well as to make Clinica available as a cloud service to make its use easier for the users.

The set of post-hoc analytic tools (statistics and machine learning) cannot cover all the potential techniques that can be applied to the extracted data. We have included what we believe are the most commonly used approaches so that they can be performed in a user-friendly way. Nevertheless, users may need additional analytical flexibility, for instance for applying other statistical models (e.g. mixed-effect models, survival analysis…) or machine learning tools. The standardized organization of the outputs as well as the presence of tools for generating a single tsv file from a hierarchy of outputs should facilitate the subsequent use of other statistical or machine learning packages.

Clinica currently has the following additional limitations. First, it currently lacks pipelines for several important functional neuroimaging modalities (functional MRI, arterial spin labelling). Second, we will improve the enforcement of reproducibility by adding traceability features. Moreover, QC of processed data is currently done using standard image viewers which is clearly suboptimal. We plan to add more advanced QC of outputs in the spirit for instance of MindControl (Keshavan et al., 2018). This is important in order to ease the QC of large datasets and to enforce the good practice of systematic QC among the users. Implementing an integrated QC system is among our priorities for the development of Clinica. The aim is to provide a visual dashboard that would allow the user to: 1) easily control which pipelines have been executed and whether they exited without error; 2) systematically review snapshots of the major outputs of each pipeline (together with typical examples of how a correct output should look like); 3) flag incorrect outputs so that they can be excluded from further statistical analysis. Another limitation is that, in order to use previously built templates, they currently need to be manually copied into the CAPS directory which is clearly suboptimal. This will be changed in the near future. Finally, new longitudinal analysis pipelines (beyond those already present for FreeSurfer and PET surface processing) will be developed.

In conclusion, Clinica is an open-source software platform that provides a comprehensive set of processing pipelines for different neuroimaging modalities. It builds upon existing standards and software tools developed by the community. It can make clinical neuroimaging studies easier to perform and more reproducible.

## 8    Conflict of Interest

The authors declare that the research was conducted in the absence of any commercial or financial relationships that could be construed as a potential conflict of interest.

## 9    Author Contributions

Study concepts/study design, all authors; manuscript drafting or manuscript revision for important intellectual content, all authors; approval of final version of submitted manuscript, all authors; literature research, AR, NB, OC; and manuscript editing, all authors.

## 10    Funding

The research leading to these results has received funding from the programs "Investissements d'avenir" ANR-10-IAIHU-06 (Agence Nationale de la Recherche-10-IA Institut Hospitalo-Universitaire-6), ANR-11-IDEX-004 (Agence Nationale de la Recherche-11- Initiative d'Excellence-004, project LearnPETMR number SU-16-R-EMR-16) and ANR-19-P3IA-0001 (PRAIRIE 3IA Institute), from the European Union H2020 program (project EuroPOND, grant number 666992, project HBP SGA1 grant number 720270), from the joint NSF/NIH/ANR program "Collaborative Research in Computational Neuroscience" (project HIPLAY7, grant number ANR-16-NEUC-0001-01), from Agence Nationale de la Recherche (project PREVDEMALS, grant number ANR-14-CE15-0016-07), from the ICM Big Brain Theory Program (project DYNAMO), from the Inria Project Lab Program (project Neuromarkers), from the European Research Council (to Dr Durrleman project LEASP, grant number 678304), and from the "Contrat d'Interface Local" program (to Dr Colliot) from Assistance Publique-Hôpitaux de Paris (AP-HP). N.B. received funding from the People Programme (Marie Curie Actions) of the European Union's Seventh Framework Programme (FP7/2007-2013) under REA grant agreement no. PCOFUND-GA-2013-609102, through the PRESTIGE programme coordinated by Campus France.